\documentclass{svjour3}
\usepackage{epsfig}
\usepackage{graphicx}
\usepackage{amssymb}
\usepackage{bm}
\usepackage{amsmath}

\def\beq{\begin{equation}}
\def\eeq{\end{equation}} 
\def\br{\begin{eqnarray}}
\def\er{\end{eqnarray}}
\def\benu{\begin{enumerate}}
\def\eenu{\end{enumerate}}
\def\nn{\nonumber}

\begin{document}

\title{Competing with oneself: Introducing self-interaction \\ in a model of competitive learning}

\author{Gaurang Mahajan    \and
        Anita Mehta}

\institute{Gaurang Mahajan \at 
           S. N. Bose National Centre for Basic Sciences, Block JD, Sector III, \\ Salt Lake, Kolkata 700 098, India \\
\email{gaurang@iucaa.ernet.in} 
\and 
Anita Mehta \at 
S. N. Bose National Centre for Basic Sciences, Block JD, Sector III, \\ Salt Lake, Kolkata 700 098, India \\
\email{anita@bose.res.in}
}

\date{Received: date / Accepted: date}

\maketitle
\begin{abstract}
A competitive learning model was introduced in Mehta and Luck (1999) (Phys Rev E 60, 5: 5218-5230), in which the learning is outcome-related. Every individual chooses between a pair of existing strategies or types, guided by a combination of two factors: tendency to conform to the local majority, \emph{and} a preference for the type with higher perceived success \emph{among its neighbors}, based on their relative outcomes. Here, an extension of the \emph{interfacial model} of Mehta and Luck (1999) is proposed, in which individuals additionally take into account their \emph{own} outcomes in arriving at their outcome-based choices. Three possible update rules for handling \emph{bulk} sites are considered. The corresponding phase diagrams, obtained at \emph{coexistence}, show systematic departures from the original interfacial model. Possible relationships of these variants with the \emph{cooperative model} of Mehta and Luck (1999) are also touched upon.
\end{abstract}

\section{Introduction}

The application of tools and concepts of statistical physics in the modeling of social behavior has received considerable attention in recent times (for a recent review see, e.g., Castellano et al. 2009). There are instances, in social and biological contexts, of regularities emerging on the collective scale even when the interactions of any given individual are confined to a very small subset of the population, and these examples share similarities with cooperative phenomena in other physical systems. Such social interactions may well take the form of \emph{learning}, describable in very general terms as mimicking, or adopting, the behavior of other individuals. 
From the perspective of game theory, this would be seen as an adoption of a particular \emph{strategy}, whose result may or may not be associated with a favorable \emph{outcome}.  It is quite reasonable to expect that the effectiveness of a strategy in yielding favorable outcomes would influence how likely it is to persist, and spread through the population. 

Against the backdrop of the above ideas, a model of strategic learning was introduced in Mehta and Luck (1999) which considers a population of interacting agents on a regular lattice, with one of two possible strategies, or types, being associated with each. At every time step of the dynamics, each agent independently makes a choice to either remain unchanged in type or switch to the other type, based on two elementary rules: a \emph{majority-based} updating, reflecting every agent's tendency to align with the local neighborhood, followed by a \emph{performance-based} updating depending on the [random] outcomes of \emph{its neighbors} of both types in some `game'. An outcome is categorized as being either a success (with a fixed probability $p_{\pm}$ for each type) or a failure; every individual revises its choice, adopting the more successful of the two types among its neighbors. This model, in its two versions, interfacial and cooperative, was investigated in depth in Mehta and Luck (1999) at \emph{coexistence}, i.e., when $p_+$ = $p_-$ = $p$. Using an analytical pair approximation as well as numerical simulation, the corresponding phase diagrams were constructed. In the case of the interfacial model, the phase diagram was found to possess a regime of disorder for intermediate values of the performance parameter $p$, flanked by frozen phases on either side; a correspondence was exhibited between regimes close to $p=0$ and $p=1$.   

In what follows, a variant of the interfacial learning model is considered with the addition of one extra ingredient: the introduction of self-coupling in the performance-based updation rule. This is a natural extension of the previous work on this model, and would perhaps bring it closer to capturing the dynamics of a group of individuals in contexts where some form of learning by observation may be involved. In certain situations, choices made by individuals between available alternatives might be based not only on which of the alternatives is more widespread in the population, but also on which of them is \emph{perceived} to be more `successful' at doing whatever they are supposed to do. If agents are going to decide between sticking to their own type and changing to the other type depending on the relative outcomes, then it is reasonable to suppose that every agent would observe and factor in the result of its \emph{own} chosen course \emph{as well as} the realizations of its neighbors, in making these choices. As plausible real-world examples which such a model might approximate, one can think of consumers being influenced by the experiences of friends in deciding which brand of product to buy, or farmers opting for a particular agricultural practice/crop variety on the basis of their neighbors' successes, or even perhaps, the competition that may have existed between prevalent hunting strategies in pre-historic social groups. We mention here that the kernel of the above idea has appeared earlier in some work on the learning-based diffusion of technology (Chatterjee and Xu 2004). However, that study was limited to a one-dimensional setting, and in its details differs considerably from the model to be analyzed here.

This paper proceeds as follows: In the next section, the modified version of the learning model is introduced, and three possible ways of incorporating the self-coupling are described. Subsequently, the results of the numerical analysis of these models on a two-dimensional square lattice are presented in Sec. III. This is followed in Sec. IV by a recapitulation and some discussion of the relationship of the present model with the cooperative model of Mehta and Luck (1999). We conclude by putting our work in the context of previous models of learning in Sec. V.

\section{The Model}

In the standard interfacial model as defined in Mehta and Luck (1999), individuals are located at the sites of a regular lattice with coordination number $z$ (equal to 4 for a square lattice in two dimensions) and every individual is associated with an \emph{efficiency} $\eta_i$, which can take one of two possible values: +1 or -1. These alternatives will be referred to as `$+$' types and `$-$' types respectively. The dynamical evolution proceeds via a two-step process: a majority-based updation (step 1) followed by a performance-based rule (steps 2 and 3). In the latter step, every \emph{interfacial} site compares the `successes' of its neighbors of both types (more specifically, the ratios $r_+$ and $r_-$ representing the \emph{fractions} of successful neighbors of each type) averaged over some observation period, and based on their relative values, decides to either convert to the other type or remain unchanged. \emph{Bulk} sites, having all neighbors of a single type, are left untouched by the outcome-based rule in this model, as its name suggests.

Mehta and Luck (1999) also introduce a \emph{cooperative} variant of the above model. Very briefly, the cooperative model involves the same update rules as the interfacial model for interfacial sites, but introduces the additional possibility that, in the event of a majority of the neighbors of a bulk site failing, they are \emph{persuaded} to convert to the other type \emph{along with} the central site. (If the central site is \emph{already} of the other type, then of course it stays the same.)

Here, it will be assumed that step 1 of the interfacial model remains unaffected, and as before, the efficiencies are updated according to the following majority rule:
\br 
\eta_i (t+1) &=& +1 ~~\textrm{if}~~ h_i(t)>0, \nn \\
\eta_i (t+1) &=& +1 ~\textrm{or}~ -1 ~~\textrm{w.p.}~~ 1/2 ~~\textrm{if}~~ h_i(t)=0, \nn \\
\eta_i (t+1) &=& -1 ~~\textrm{if}~~ h_i(t)<0. \nn \\
\er
Following Mehta and Luck (1999), $h_i(t)$ denotes the \emph{local} field of site $i$, defined as the sum of efficiencies of its neighbors. 

It is only the second step which will be altered by now taking every site's \emph{own} outcome into account too: In addition to its neighbors on the lattice, the realizations of the central site are also included in the determination of the corresponding ratios $r_+$ and $r_-$. A departure from the properties of the original model may thus be expected.

For a start, the inclusion of outcomes of one extra site in the performance-based updation rule changes the form of the \emph{transition probabilities} $w_{\pm}(h_i)$. These encapsulate the net effect of steps 2 and 3 in the limit when the time scale on which individuals change \emph{type} is much slower than the rate at which the \emph{outcomes} of individuals are updated. In the symmetric case ($p_+ = p_-$), these probabilities are now given by
\br
\eta_i (t) = +1 ~&\rightarrow&~ \eta_i (t+1) = +1 ~~\textrm{w.p.}~~ w_+ (h_i), \nn \\
\eta_i (t) = -1 ~&\rightarrow&~ \eta_i (t+1) = +1 ~~\textrm{w.p.}~~ w_- (h_i)  \nn
\er
where
\begin{eqnarray}
w_+ (+4) &=& 1 \nn \\
w_- (+4) &=& 1 - p - (1-p)^5 \nn \\
w_+ (+2) &=& p^5 + 1 - p \nn \\
w_- (+2) &=& 1 - (1-p)^5 - p^2 - 2 p (1-p)^3 (1+2p) \nn \\
w_+ (0) &=& 1 - p^2 (1-p^3) - 2 p (1-p)^3 (1+2p) \nn \\
w_- (0) &=& 1 - p^5 - (1-p)^5 - 3p(1-p)^4 \nn \\
&&- (1-p^2)(3p^2 - 2p^3) \nn \\
w_+ (-2) &=& 1 - 3p(1-p)^4 - (1-p^2)(3p^2 - 2p^3) \nn \\
w_- (-2) &=& p - p^5 \nn \\
w_+ (-4) &=& p +(1-p)^5 \nn \\
w_- (-4) &=& 0.   \label{scm1}
\end{eqnarray}
Here, we have followed the convention of Mehta and Luck (1999) and continue to treat the probabilities as functions of the local field $h_i$, even though their determination would now involve input coming from the central site as well. These probabilities satisfy the condition $w_+(+h_i)+w_-(-h_i)=1$, which follows from symmetry under a global reversal of signs. (The derivation of the above expressions is summarized in the Appendix.) The extra feature being introduced here, compared with the interfacial model, is the possibility of conversion even when a site might be surrounded by all neighbors of a single type: For example, if $\eta_i (t) = -1$ and is surrounded by all neighbors of the `$+$' type, there is a non-zero probability, given by $w_- (+4)$, that it will switch to the `$+$' state at step 3 (this cannot happen in the interfacial model of Mehta and Luck (1999)). Loosely speaking, the self-coupling increases the effective noise in the system by adding to the randomness in the dynamics of the outcome-based update rule -- `noise' here referring to changes which do not obey a simple majority rule. A widening of the disordered paramagnetic phase may thus be expected. 

The above modification (referred to hereafter as SCM1 $\equiv$ Self-Coupled Model 1), however, does not introduce the potential for changes in a \emph{sea} of individuals belonging to the same type. (To avoid ambiguity, it is clarified here that our notion of a `sea', in which the central site \emph{as well as} its neighbors are identical, is distinct from `bulk', by which we only mean that all the neighbors of a given site have the same type, but are not necessarily identical to it.) In fact, the probabilities $w_- (+4)$ and $w_+ (-4)$ only act to \emph{increase} the order locally. One could therefore think of an alternative modification of the following kind: If a site is surrounded by neighbors all of which are of its own type, then it will decide to flip to the other type if, and only if, \emph{all} of them (including itself) fail. In this particular situation, the central site is unable to make a comparison between the two types since all its neighbors as well as itself have the same type. This rule can thus be thought of as the central site deciding to take a risk and changing to the other type \emph{only} in the worst-case scenario of total failure of its own type (i.e. itself \emph{and} all its neighbors). It may be noted that this ``all-or-nothing'' rule is different from step 3 of the \emph{cooperative} model of Mehta and Luck (1999) in two respects: One, only the central site flips here, so there is no \emph{persuasion} of the neighbors to convert; and secondly, this is not a `majority-based' rule, in the sense that the central site flips \emph{only if} \emph{all} $z+1$ sites fail, not just if more than half of them fail (which is the case for the cooperative model). 

This second version of self-coupling modifies only the probabilities $w_{+} (4)$ and $w_- (-4)$, which are now given by
\br
w_+ (+4) &=& 1 - (1-p)^5 \nn \\
w_- (-4) &=& (1-p)^5,  \label{scm2}
\er
and this variant will be referred to as SCM2. All other transition probabilities remain the same as in the SCM1 model. One can immediately see from Eq. (\ref{scm2}) that SCM2 introduces the possibility of changes in a sea of the same type; this effect is expected to increase as site failures become more probable, which would happen at sufficiently small values of $p$. This conversion of the central site to the other type could `propagate' in the course of the updating according to step 3, before the next round of step 1 (the majority rule) has a chance to `reset' it back again.

Taking the above reasoning a step further, one can possibly consider a third variant (SCM3), which implements a majority-based rule at step 3: If a site as well as all its neighbors are of the same type, the central site decides to flip to the other type if a majority of them (which on a square lattice would be equal to or greater than 3 out of 5) fail. The transition probabilities for sites in a sea now assume the form
\br
w_+ (+4) &=& 1 - (1-p)^5 - 5p(1-p)^4 - 10 p^2 (1-p)^3, \nn \\
w_- (-4) &=& (1-p)^5 + 5p(1-p)^4 + 10 p^2 (1-p)^3  \label{scm3}
\er
with the rest of the probabilities remaining the same as in SCM1 and SCM2. At $p=0$, the expressions in Eq. (\ref{scm3}) obviously coincide with the corresponding probabilities for SCM2 in Eq. (\ref{scm2}), since here \emph{all} the sites fail in any case. However, for any non-zero value of $p$, SCM3 introduces a greater likelihood of site conversions inside ordered domains.

There are thus two kinds of effective noise being introduced by the dynamics of the outcome-based update rule: the \emph{interfacial} noise which affects sites having neighbors of \emph{both} types, and \emph{bulk} noise which is associated with the updating of sites all whose neighbors are of one type. Referring to the models defined here, SCM1 contains only interfacial noise, but in the SCM2 and SCM3 models, bulk noise is also present. 

An attempt to understand the resulting dynamical phases of the above models will be made in the following sections. Two-dimensional square lattices of size $N \times N$ are considered and periodic boundary conditions (with the identification of opposite edges) are imposed. In order to quantify the nature of the collective dynamics and identify various phases, the averages of two quantities will be looked at. These are the \emph{magnetization} $M(t)$, defined as the mean efficiency over the whole lattice, and the \emph{energy} $E(t)$, given by the fraction of \emph{disparate}, or active, bonds (i.e. those which connect sites with unequal efficiencies):
\br
M &=& \frac{1}{N^2} \sum_i \eta_i \\
E &=& \frac{1}{4N^2}\sum_{(ij)} (1-\eta_i \eta_j).
\er 
The second summation above is taken over all possible pairs of nearest-neighbor sites. With this definition, the energy is zero when full consensus is reached in the population (corresponding to $M = \pm 1$), and maximized (equal to +1) in an \emph{antiferromagnetic} state. 

At a qualitative level, an idea of the behavior of these models may be had by making a correspondence with a generalized two-parameter stochastic model studied in Drouffe and Godr\`{e}che (1999) and Oliveira et al. (1993), as was done in Mehta and Luck (1999). The transition probabilities in this class of nonequilibrium spin models are parametrized by two variables $p_1$ and $p_2$, interpreted respectively as measures of interfacial and bulk noise. Specific points in the $p_1-p_2$ plane correspond to particular cases, like the zero-temperature Ising-Glauber model ($p_1 = p_2 = 1$), and the voter model ($p_1 = 3/4, p_2 = 1$) (Holley and Liggett 1975; Frachebourg and Krapivsky 1996; Dornic et al. 2001). The analysis in Drouffe and Godr\`{e}che (1999) and Oliveira et al. (1993) showed that the phase diagram is divided in a ferromagnetic region near the $p_1 = 1, p_2 = 1$ point and a disordered paramagnetic phase, which are separated by a line of \emph{continuous} phase transitions terminating at one end at the voter point. In Mehta and Luck (1999), the interfacial model with no bulk noise was associated with the $p_2 = 1$ line, and the critical point was shown to have the character of a \emph{first-order} transition belonging to the same universality class as the voter model, which shows criticality in two dimensions. 

Considered in the above context, each self-coupled model, in a rough sense, may be represented by a curve parametrized by $p$ in the $p_1-p_2$ plane. These curves would all converge at the zero-temperature Ising point when $p=1$, and cross the critical line into the disordered phase at different points (equivalently, at different values of $p < 1$). In SCM1, bulk noise (taken to mean a non-zero probability for a site surrounded by all neighbors of its type to flip to the opposite type) is absent. Thus, this version would follow the $p_2 = 1$ line just as the interfacial model, going from fully ordered ferromagnetic to paramagnetic behavior; the corresponding critical point may be expected to be voter-like with a discontinuity in the value of order parameter. As for SCM2 and SCM3, there is also the presence of bulk noise in addition to interfacial noise, implying that the order-disorder transition in these models would be of the continuous type. Moreover, while the interfacial noise is identical in all the three models -- allowing a direct correspondence to be made with $p$ -- there is an overall enhancement of bulk noise in SCM3 compared to SCM2. Referring back to the phase diagram and the critical line of the generalized model suggests that the critical point of SCM2 would occur at higher interfacial noise, i.e. at a smaller value of $p$, relative to SCM3 (this is also anticipated on general grounds). These expectations will be borne out by the analysis in the next section.

\section{Numerical results}

The time-evolution of the lattice dynamics is implemented numerically. In every run, the population is assumed to start out in a random initial configuration of efficiencies, in which each site is independently chosen to be of the `$+$' type or `$-$' type with equal probability. As for the updating procedure to be employed, different options exist. A commonly used scheme in the literature for problems of this type (which, however, normally involve only a single-step update rule) is \emph{random} sequential updating (see, e.g., Christensen and Moloney 2005; Landau and Binder 2005) in which one site is picked at random at every moment and its state updated; this means that any given site in the lattice gets updated once every $N^2$ time steps only \emph{on an average}. But here, following Mehta and Luck (1999), the updating will be carried out in an \emph{ordered} sequential manner, which consists of ``sweeping'' through the lattice in a regular fashion (left to right and top to bottom, say) so that adjacent sites along any one direction get updated in succession. One complete round of updating is implemented by first sweeping through the lattice updating sites according to step 1, followed by a sweep of updating based on the probabilities $w_{\pm}$. These two sweeps together will be taken to constitute one ``time step'' of the dynamics.

\begin{figure}[t]
\begin{center}
\includegraphics[scale=0.55]{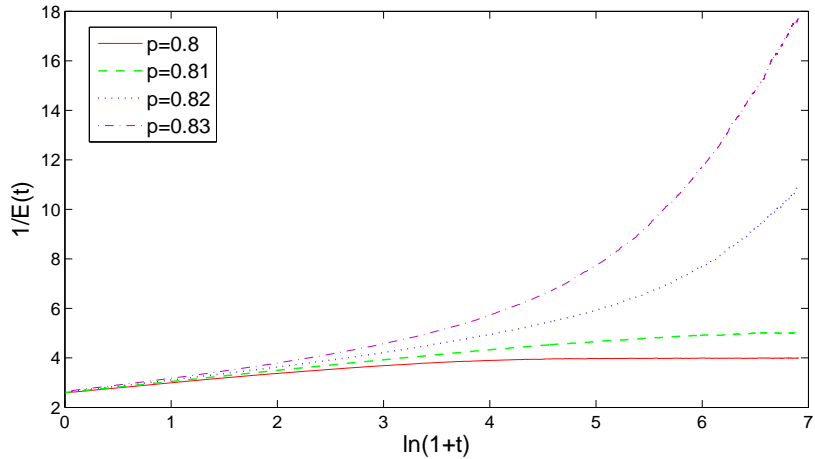}
\caption{\label{fig:scm1a} Plots of the inverse of the mean energy $E(t)$ as function of time in the vicinity of $p=0.8$ for the self-coupled model 1 (SCM1). System size is $500 \times 500$.}
\end{center}
\end{figure}

\begin{figure}[h]
\begin{center}
\includegraphics[scale=0.55]{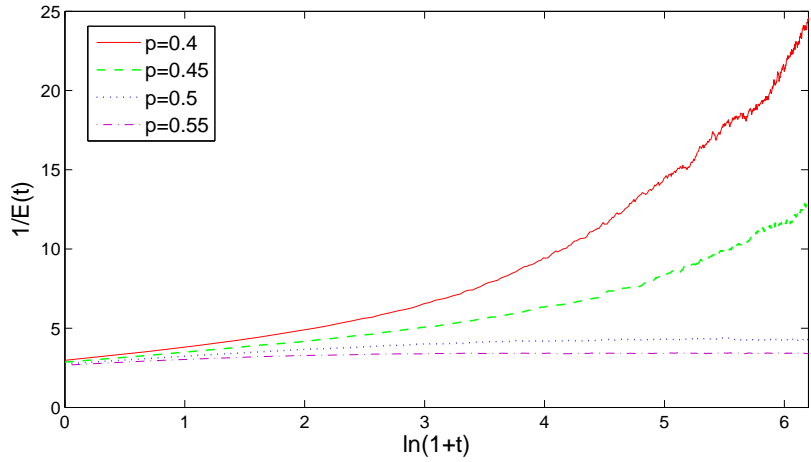}
\caption{\label{fig:scm1b} Evolution of the mean energy with time for $p$ values around 0.5 for the SCM1 model. System size is $500 \times 500$.}
\end{center}
\end{figure}

We first consider the SCM1 model. Figs.~\ref{fig:scm1a} and \ref{fig:scm1b} show the evolution of the average energy with time for different values of $p$, for a square lattice of size 500 $\times$ 500 over $T=1,000$ time steps. Each curve is an average over 100 runs (with different starting seeds). The narrow range of $p$ values displayed in Fig.~\ref{fig:scm1a} has been chosen to bring out the crossover occurring in the behavior of the system between $p = 0.81$ and $0.82$, which is reflected in the change in the asymptotic nature of the curves. The occurrence of a similar transition near $p \approx 0.5$ can be inferred from Fig. \ref{fig:scm1b}. 

\begin{figure}[t]
\begin{center}
\includegraphics[scale=0.55]{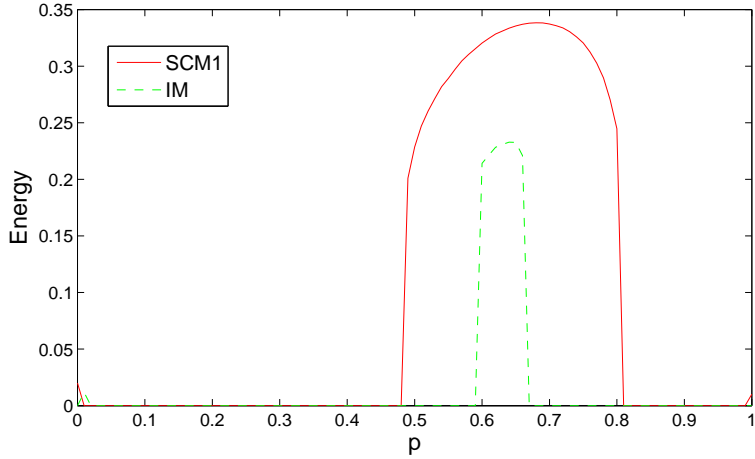}
\caption{\label{fig:scm1E} Comparison of mean energy vs. $p$ for the interfacial model (IM) and self-coupled model 1 (SCM1). Lattice size is $100 \times 100$.}
\end{center}
\end{figure}

\begin{figure}[t]
\begin{center}
\includegraphics[scale=0.55]{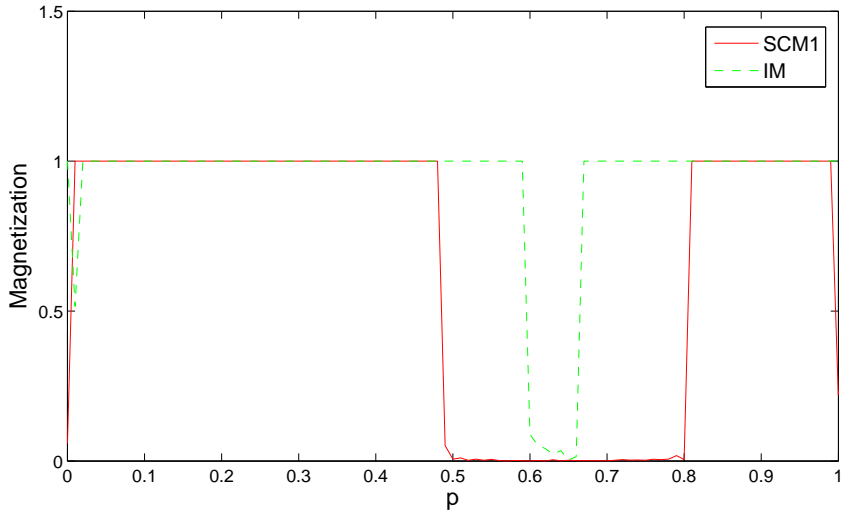}
\caption{\label{fig:scm1M} Comparison of mean magnetization vs. $p$ for the interfacial model (IM) and self-coupled model 1 (SCM1). (Lattice size is $100 \times 100$.)}
\end{center}
\end{figure}

This behavior shows up in the plots in Figs. \ref{fig:scm1E} and \ref{fig:scm1M} of the average energy and average magnetization as functions of the probability $p$; the corresponding results for the interfacial model are also shown for comparison. A square lattice of size $100 \times 100$ has been used here (and in the subsequent results); this system size is found to be large enough to give reliable averages, and the deviations of its predictions from larger lattices are insignificant. For obtaining the data points, runs lasting $T=50,000$ time steps each have been carried out with averaging being done over the last 10,000 steps. These graphs clearly indicate a paramagnetic phase with zero net magnetization occurring between $p \approx 0.5-0.8$, with macroscopic order in the remainder of $p$-space. The transition from disorder to fully ferromagnetic is quite sharp, consistent with the expectation of a first-order phase transition for this model. These results bear a resemblance to the interfacial model in Mehta and Luck (1999); the similarity is expected, given that in both models it is only the interfacial noise which is present. However, the interval of disorder in SCM1 is almost doubled in comparison with the interfacial model, which may be attributed to an overall enhancement of the interfacial noise due to the self-coupling.

The nature of the ordered phases was probed a little further by looking at the time series of individual runs (starting from random initial configurations) with different choices for $p$. In all the runs carried out, the system was observed to always get completely ordered when evolution was allowed to proceed for sufficiently long times. Also, since bulk noise is absent in this particular model of self-coupling -- that is, individuals cannot break out of a sea -- the fully ordered state is absorbing. At the two end points $p=0$ and $p=1$, however, it is found that the system can also end up in frozen \emph{stripe} states which are only partly ordered and remain stable over time. At these end points, where all sites succeed or all of them fail with certainty, the outcome-based rule in step 3 becomes irrelevant and evolution is driven solely by the zero-temperature Ising update rule of step 1. The frozen states seen here are consistent with earlier results for quenching and phase ordering in zero-temperature Ising models (Spirin et al. 2001a, b; Barros et al. 2009). Based on this, it is conceivable that sufficiently close to $p=0$ and $p=1$ where the interfacial noise introduced by step 3 would be very weak, the system may not get fully ordered for some initial seeds. It could, for example, remain trapped in partially ordered configurations indefinitely; in such cases, while the interfacial noise might induce some activity at the boundaries it could be too weak to induce transitions to the fully ferromagnetic state. This possibility has not, however, been looked into in detail in the present work.    

\begin{figure}[t]
\begin{center}
\includegraphics[scale=0.55]{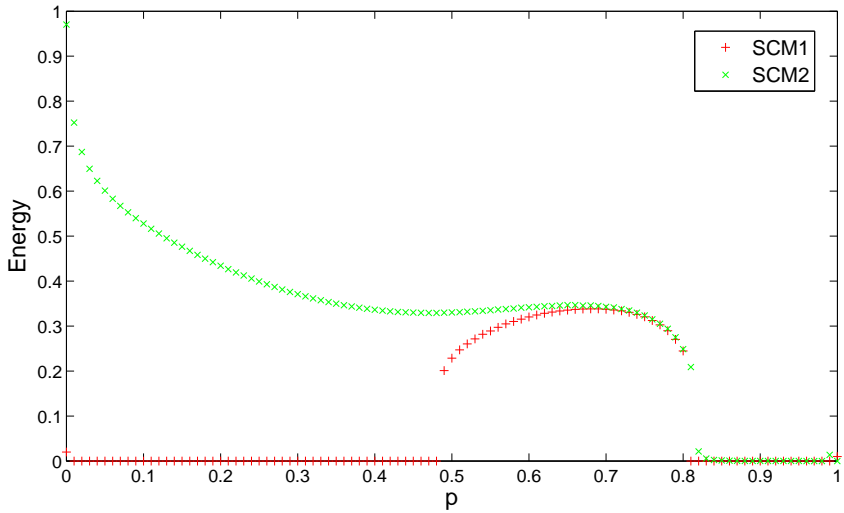}
\caption{\label{fig:scm2E} Comparison of mean energy vs $p$ for the two self-coupled models SCM1 and SCM2, for system size of $100 \times 100$.}
\end{center}
\end{figure}

\begin{figure}
\begin{center}
\includegraphics[scale=0.55]{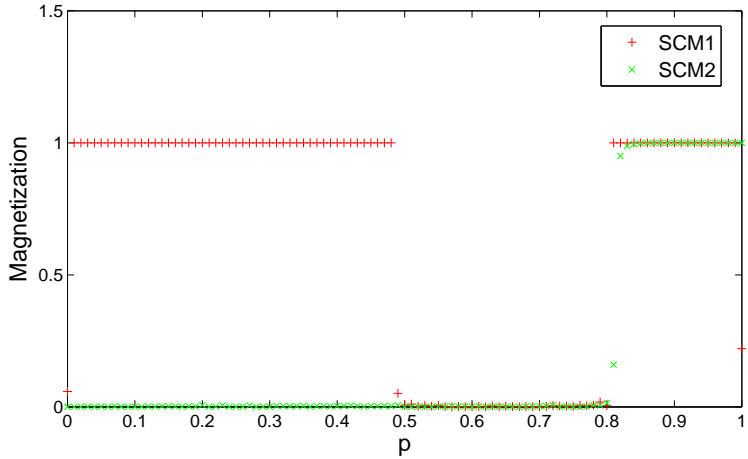}
\caption{\label{fig:scm2M} Comparison of mean magnetization vs $p$ for the self-coupled models SCM1 and SCM2 (again for a 100 $\times$ 100 lattice).}
\end{center}
\end{figure}

\begin{figure}[t]
\begin{center}
\includegraphics[scale=0.55]{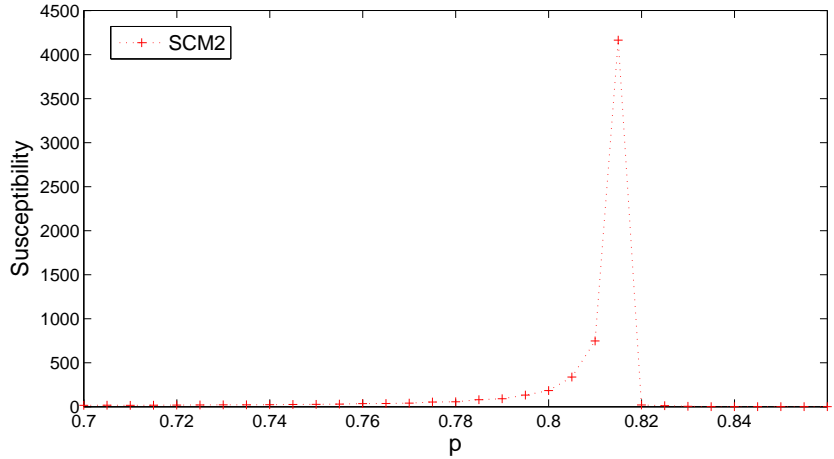}
\caption{\label{fig:scm2_sus} The susceptibility, defined in terms of the variance of $M$, as a function of $p$ for the SCM2 model (system size is 100 $\times$ 100). The capped spike occurs at the disorder-order transition.} 
\end{center}
\end{figure}

The second version with self-coupling, SCM2, is considered next. Figs. \ref{fig:scm2E} and \ref{fig:scm2M} are plots of the mean energy and mean magnetization as functions of $p$ displayed along with the corresponding curves for SCM1 (again for a 100$\times$100 lattice with every run lasting $T=50,000$ time steps and averaging being done over the last $10,000$ steps). Visually, from the plot of the magnetization, only one clear-cut transition can be made out, and it is located very near to the critical point of SCM1. But recalling the analogy with the generalized spin model of Drouffe and Godr\`{e}che (1999) and Oliveira et al. (1993), the presence of bulk noise here is expected to change the nature of the transition to a continuous type. To locate the critical point, the susceptibility, defined in terms of the variance of the net magnetization as $\chi = N^2 (\langle M^2 \rangle - \langle M \rangle^2)$ (Christensen and Moloney 2005), is considered in the vicinity of the transition and is plotted in Fig. \ref{fig:scm2_sus} (the points correspond to averages over the last 20,000 time steps in runs up to $T=100,000$). From the peak in the data, which is expected at the disorder-order transition, the critical point is located at $p_c \approx 0.815$, confirming the second-order nature of this transition. 

\begin{figure}[t]
\begin{center}
\includegraphics[scale=0.55]{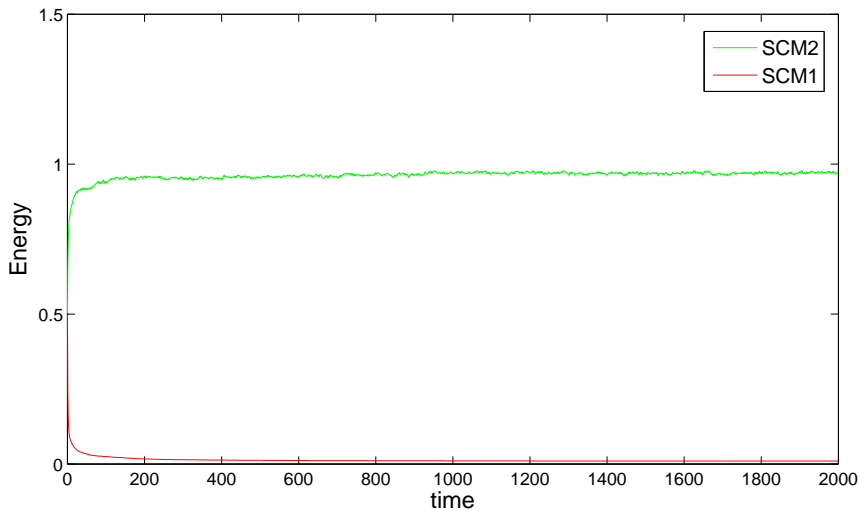}
\caption{\label{fig:scm2runE} Comparison of time evolution of energy over one run at $p=0$ for the two self-coupled models SCM1 and SCM2 (for a 100 $\times$ 100 lattice).}
\end{center}
\end{figure}

\begin{figure}
\begin{center}
\includegraphics[scale=0.55]{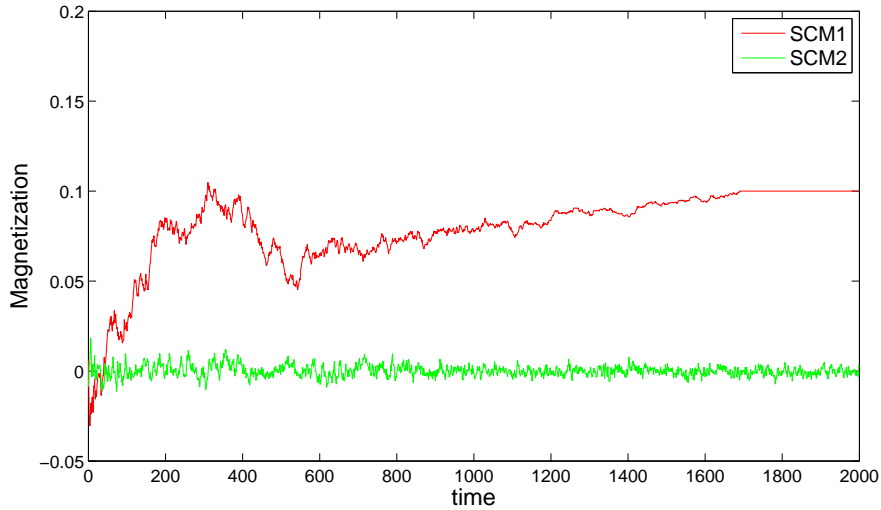}
\caption{\label{fig:scm2runM} Comparison of time evolution of the mean magnetization over one run at $p=0$ for the self-coupled models SCM1 and SCM2 (for a 100 $\times$ 100 lattice).}
\end{center}
\end{figure}

\begin{figure}[ht]
\begin{center}
\includegraphics[scale=0.55]{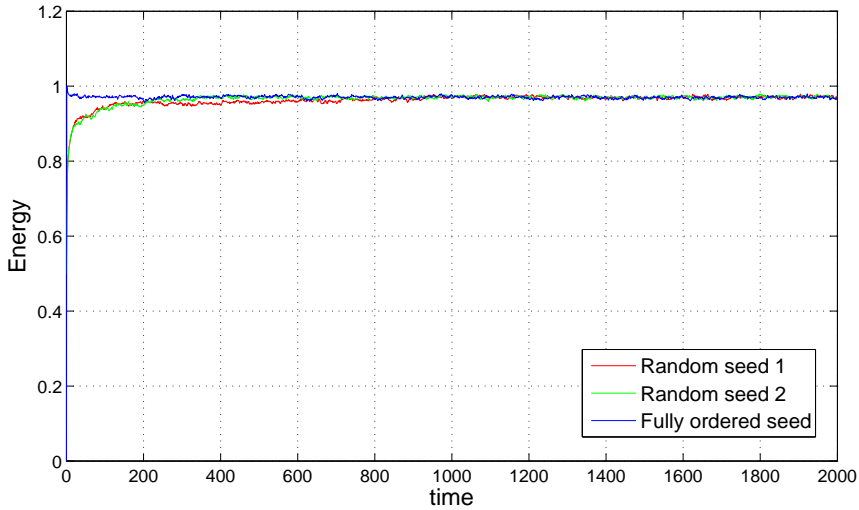}
\caption{\label{fig:scm2_E-t_comparison} Result for three runs with different seeds (including a fully ordered starting state) for the SCM2 model at $p=0$. (Lattice size = 100 $\times$ 100.)}
\end{center}
\end{figure}

Shifting attention to the other side, i.e. for smaller values of $p$, the magnetization is observed to remain nearly zero. The energy is far from minimized, and in fact, it steadily increases in the $p \to 0$ limit, suggestive of a preference for an antiferromagnetic state. However, it is found that even in the $p \to 0$ limit the proportion of active bonds does not reach unity, so the system does not appear to get antiferromagnetically aligned in its entirety. Figs. \ref{fig:scm2runE} and \ref{fig:scm2runM} show the time evolution of energy and magnetization over one run each for the SCM1 and SCM2 models (with random initial seeds) when $p=0$, at which point the difference between the two models is maximum. It can be seen that the SCM1 model gets stuck in a \emph{frozen} stripe state (at $t \approx 1,700$). In the case of SCM2, the energy asymptotes to some steady-state value slightly less than unity (with small fluctuations) and the net magnetization is zero, indicating an antiferromagnetic state. Based on several individual runs, it is confirmed that this asymptotic value of energy is independent of the initial conditions, and is a function of $p$ alone. This is illustrated in Fig. \ref{fig:scm2_E-t_comparison} with three different starting states, including two random seeds and the fully ordered state (which after a single time step converts to antiferromagnetic).   

A glimpse at the snapshot of the lattice at $p = 0$ (taken after a sufficiently long time is allowed to elapse starting from a random initial seed) shows that an \emph{almost} completely antiferromagnetic configuration is reached, consistent with the increase in the energy of the system towards the $p \to 0$ end. However, some patches of disorder are found to persist along the edges. We have checked that these are due to finite-size effects. Furthermore, in the antiferromagnetic domain, collective flipping of sites at every time step is observed. 

The preference for antiferromagnetic ordering seen in this model can be understood by noting that close to $p=0$ failures become increasingly likely, so that an individual surrounded by all neighbors of the opposite type would tend to \emph{retain} its type. In contrast, an individual in a sea of identical types would tend to do the exact opposite, choosing to convert because of all-round failure of its own type. Also, there would be no interfacial noise at $p=0$. The majority rule in step 1 acts to increase the local order. The bulk noise due to step 3 following it would then come into play and neighbors would tend to anti-align, thus favoring a configuration made up of active bonds all over. 

\begin{figure}[t]
\begin{center}
\includegraphics[scale=0.55]{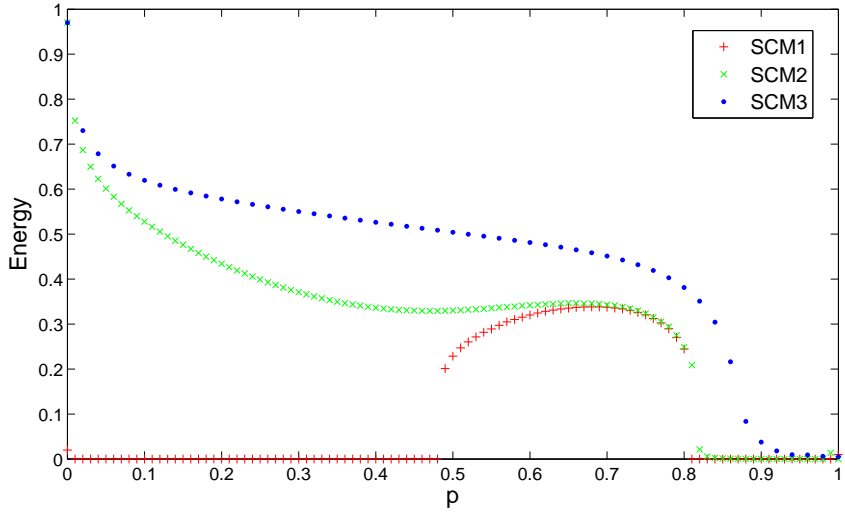}
\caption{\label{fig:scmsE} Comparison of plots for mean energy vs $p$ for all three models with self-coupling, for a 100$\times$100 lattice. The SCM3 model implements a majority rule at step 3.}
\end{center}
\end{figure}

\begin{figure}
\begin{center}
\includegraphics[scale=0.55]{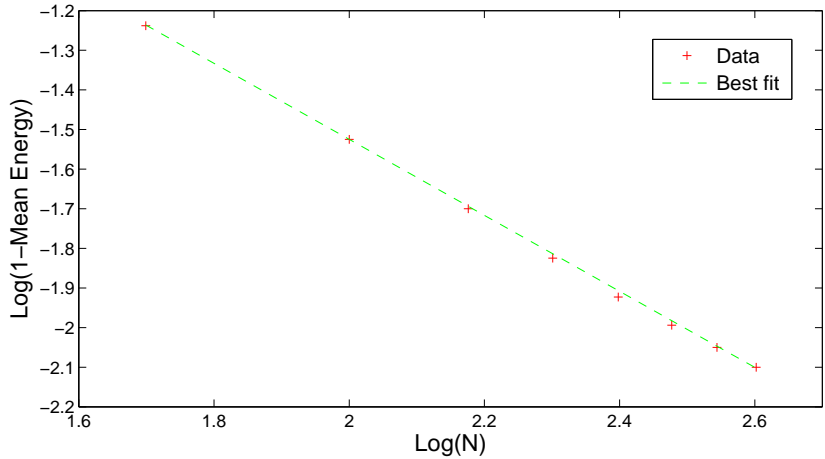}
\caption{\label{fig:scm3_E_vs_N} Log-log plot showing scaling of the mean energy at $p=0$ with lattice dimension $N$ for the SCM3 model. The best-fit straight line gives $1 - E(N) \propto N^{\delta}$ with $\delta = -0.959 \pm 0.008$.} 
\end{center}
\end{figure}

The above behavior is expected to be seen also for the third variant, SCM3, which implements a majority rule at step 3 and thus introduces even more bulk noise. Fig. \ref{fig:scmsE} displays the result for behavior of the energy, plotted against the earlier self-coupled models. SCM2 and SCM3 show qualitatively similar behavior in the small $p$ regime, but in the latter case, the antiferromagnetic alignment is comparatively stronger at any given value of $p$. (In the strict limit of $p \to 0$, however, this difference disappears as the models become essentially identical.) From the plot of Fig. \ref{fig:scmsE}, as well as from lattice snapshots taken at large times, it appears that even here, the system does not reach a fully antiferromagnetic state at $p=0$. A finite-size scaling analysis (Christensen and Moloney 2005) of this model, however, indicates that in the limit of infinite lattice size, the system energy does tend to unity. A steady reduction in the relative magnitude of the disordered area to the lattice size is implicit in this result, leading to a completely ordered antiferromagnetic state. Fig. \ref{fig:scm3_E_vs_N} shows the results for different lattice sizes ranging from $N=50$ to $N=400$ on a log-log plot, and a linear fit to the data points yields the scaling relation $1-E(N) \propto N^\delta$ with $\delta = -0.959 \pm 0.008$. 

\begin{figure}[ht]
\begin{center}
\includegraphics[scale=0.55]{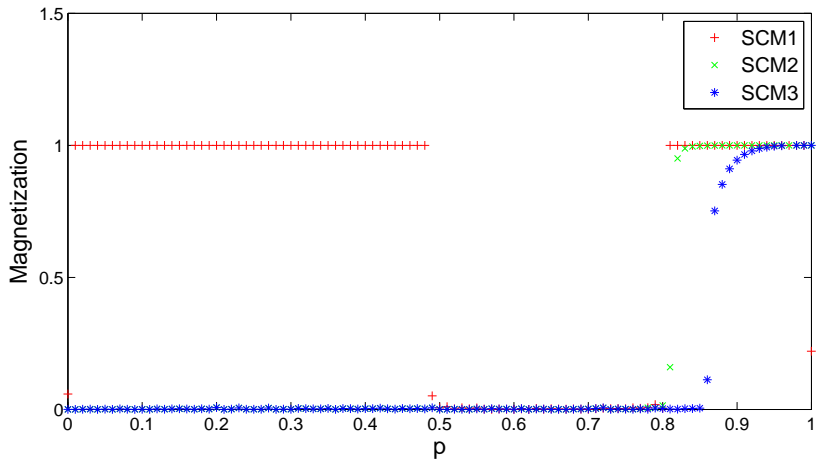}
\caption{\label{fig:scmsM} Plots of the net magnetization vs $p$ for all three models with self-coupling, for a 100$\times$100 lattice.}
\end{center}
\end{figure}

\begin{figure}
\begin{center}
\includegraphics[scale=0.55]{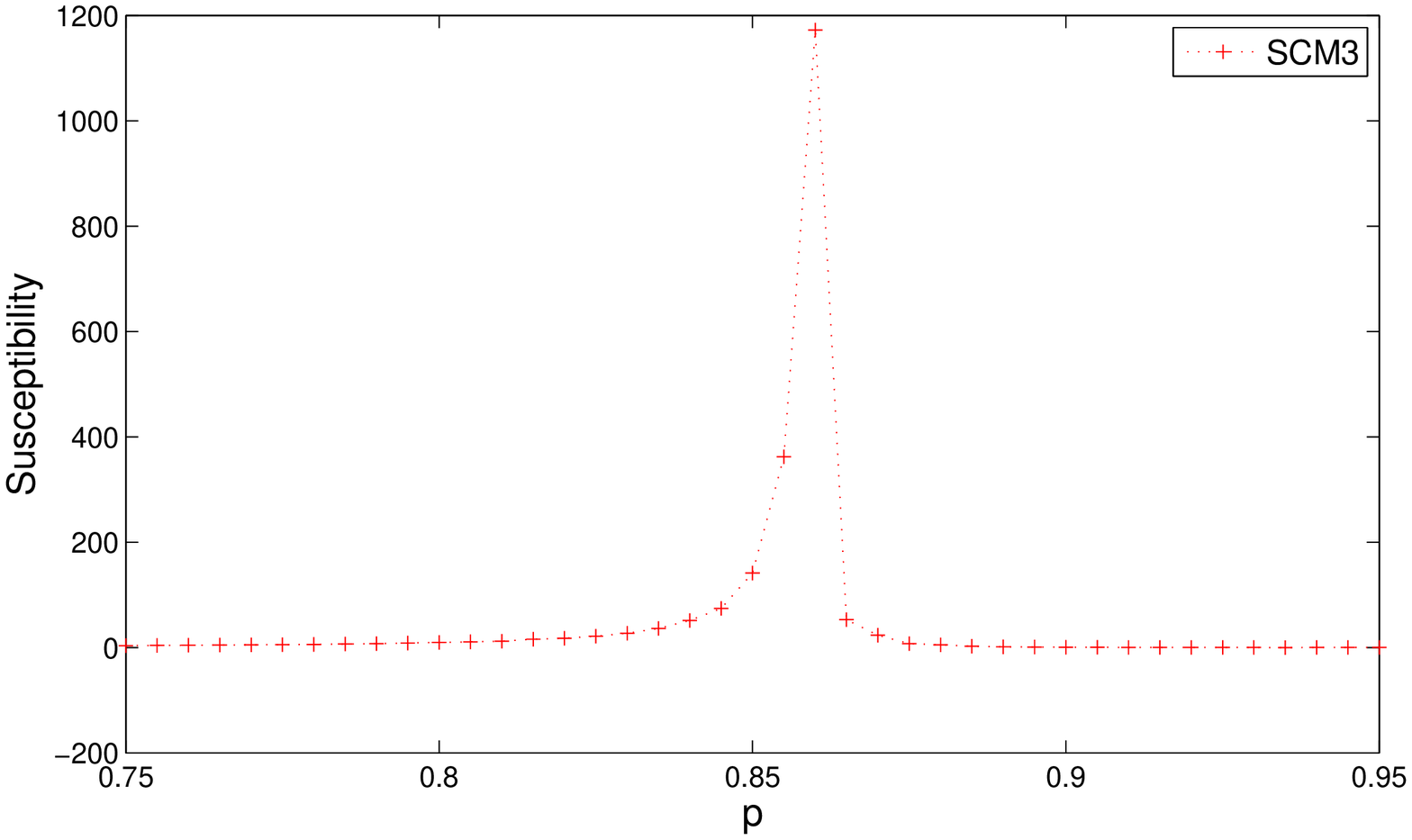}
\caption{\label{fig:scm3_sus} Dependence of susceptibility on $p$ for the SCM3 model (system size is 100 $\times$ 100). The capped spike indicates location of an order-disorder transition.} 
\end{center}
\end{figure}

Towards the other end of the phase diagram, a continuous ferromagnetic transition is found to occur beyond a critical value of $p$ which is discernible in the plot in Fig. \ref{fig:scmsM}. The corresponding data for the susceptibility as a function of $p$ in the transition region (see Fig. \ref{fig:scm3_sus}) shows a peak at $p_c = 0.86$, which is associated with the critical point. This value is higher than the corresponding value of $p_c$ for the SCM2 model. Also, the transition from disorder to ferromagnetic (i.e. complete) ordering is more gradual relative to SCM2. These differences may be put down to the comparatively higher level of bulk noise introduced by the majority rule of step 3 here (in place of the all-or-nothing rule of SCM2), which persists for larger values of $p$, and introduces greater disorder.



\section{Discussion: Quantitative analysis of results}

The model of competitive learning introduced in Mehta and Luck (1999) involves agents which make choices based only on observing their local neighborhood, and do not take into consideration the realizations of their own actions. In the preceding sections, the consequence of introducing this additional feature for the collective behavior of the population has been explored. The motivation for doing this needs hardly any justification: since agents in the original model were assumed to be sophisticated enough to learn by temporal integration of information from their local environment, it seems quite natural to \emph{also} assume some capacity for self-observation (or self-reflection), which should then have a bearing on the learning dynamics. Having said that, this particular extension is to be regarded as only one among the many possible directions in which the original model of Mehta and Luck (1999) can be taken further, and needless to say, several other possibilities remain to be explored. 

Nonetheless, the analysis carried out here suggests that the incorporation of a self-term can indeed introduce new features into the dynamics of the model. In general, the self-coupling as considered here adds to the interfacial noise, and also brings in some bulk noise. The question of which update rule to adopt when an individual happens to be at a bulk site was explored systematically with three possible choices; the corresponding phase diagrams do indeed show some significant differences. On allowing for the possibility of conversions inside a sea of individuals of the same type, as has been done in the SCM2 and SCM3 models, a preference for antiferromagnetic ordering is seen in the lattice as the $p=0$ end point is approached, in which limit failures become increasingly more frequent. An antiferromagnetic state would be interpreted as a tendency of individuals to continually maintain a position which is at odds with their local neighborhood. Additionally, introduction of bulk noise into the system changes the character of the phase transition occurring at the high-$p$ end, replacing the discontinuity in SCM1 with a continuous transition from paramagnetic to ferromagnetic behavior. This is a feature that the models SCM2 and SCM3 share in common with the \emph{cooperative model}, which was introduced in Mehta and Luck (1999). In going from SCM1 to SCM3, as the effective bulk noise being introduced into the system increases, the location of the critical point as well as the onset of full ferromagnetic ordering are pushed further towards the $p=1$ end.

It may also be pointed out that the self-coupled versions are very different from the cooperative model at the low-$p$ end of the phase diagram. The cooperative model incorporating a hard rule (Mehta and Luck 1999) predicts a phase of \emph{oscillatory coarsening} in a narrow interval near $p=0$, which is not reproduced here merely by addition of self-coupling to the original interfacial model. The existence of this time-dependent oscillatory phase is presumably related to the fact that in the cooperative model, not only can a bulk site change, but it can also persuade its neighbors to convert to the other type. This guess is supported by looking at the implementation of the cooperative model with the \emph{soft} rule, according to which \emph{only} the failed sites among its neighbors accompany the central site in converting to the other type. This version also shows oscillations near $p=0$. If one takes the soft rule a step further, and implements an even weaker version of it in which \emph{only} the central bulk site converts [provided that a majority of its neighbors fail, as before], it turns out that the oscillatory phase disappears and, in fact, gets replaced by antiferromagnetic behavior at the low-$p$ end. It may be noted that this weaker-than-soft scheme is actually quite similar to the SCM3 model (leaving aside the self-coupling of the latter). The above results thus suggest the interpretation that persuasion of neighbors to convert \emph{collectively}, a feature which is present in both the hard and soft versions of the cooperative model, but not in the other models, may be necessary to sustain the peculiar \emph{oscillatory} behavior.

\begin{figure}[t]
\begin{center}
\includegraphics[scale=0.55]{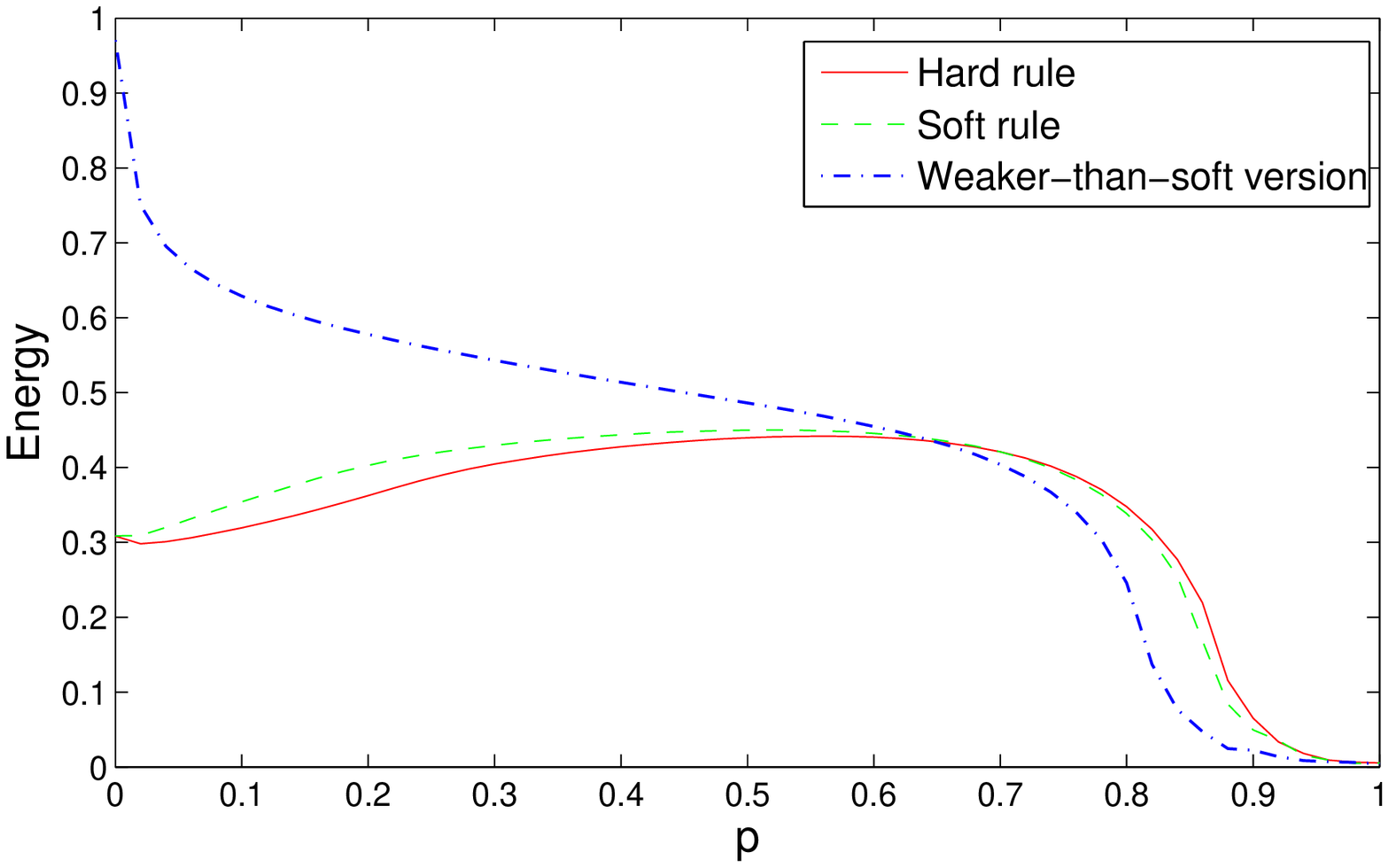}
\caption{\label{fig:cms} Comparison of mean energy vs $p$ curves for all three versions of the cooperative model. (Lattice size = 100$\times$100.)}
\end{center}
\end{figure}

\begin{figure}[h]
\begin{center}
\includegraphics[scale=0.55]{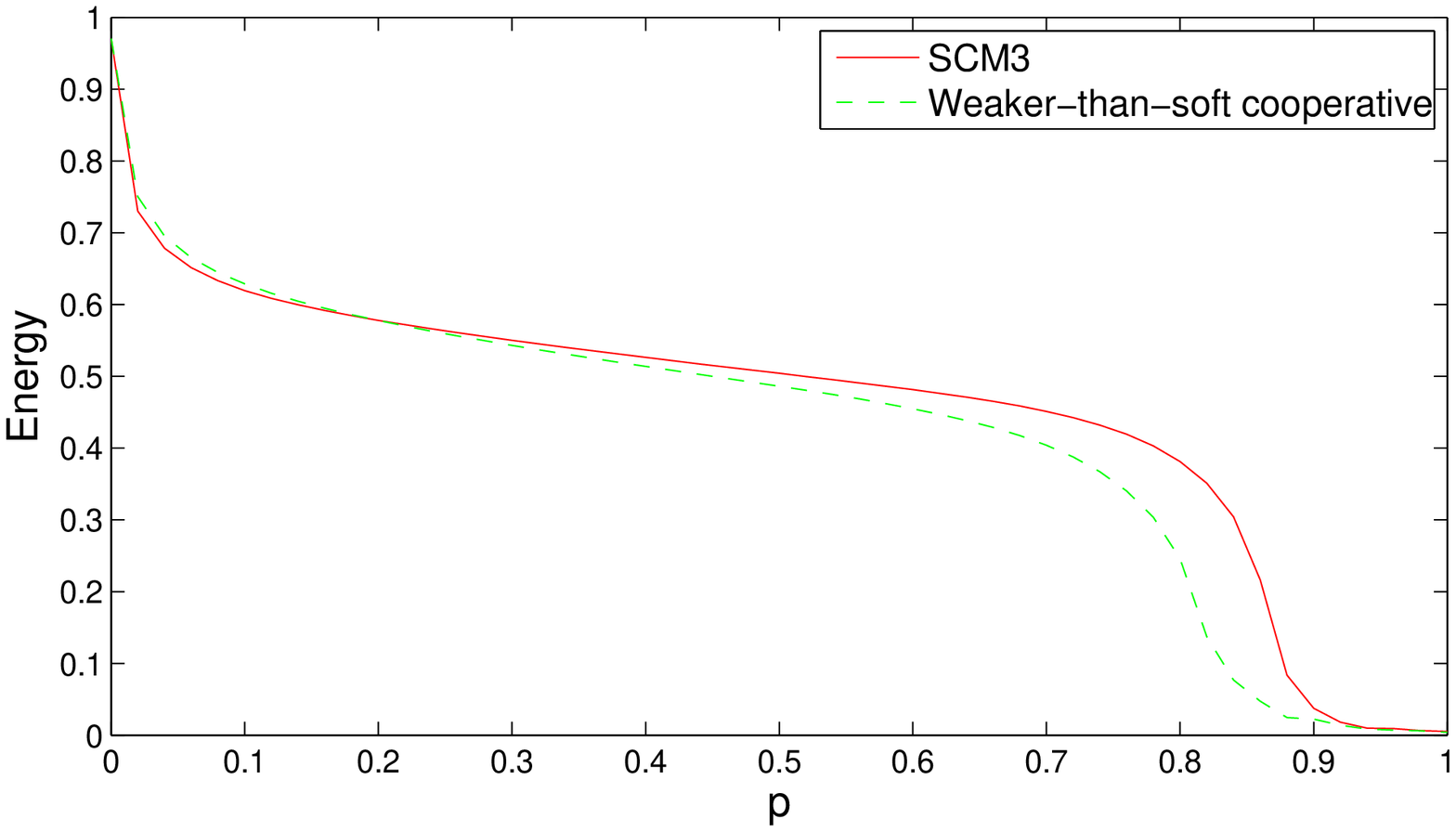}
\caption{\label{fig:cmsr2scm3} Comparison of weaker-than-soft version of the cooperative model with SCM3 (Lattice size = 100$\times$100.)}
\end{center}
\end{figure}

Going a bit further, the variation of mean energy with $p$ has been plotted for the three versions of the cooperative model in Fig. \ref{fig:cms}. For comparison, the result for the weaker-than-soft version of the cooperative model is also shown along with the corresponding curve for SCM3 in Fig. \ref{fig:cmsr2scm3}. The onset of ordering in the case of the weaker-than-soft version occurs at a smaller value of $p$ relative to the other cases. At a qualitative level, these differences are traceable to the differences in the effective `noise' being introduced by the outcome-related rule in the various models, and this may be seen as follows: At $p=1$, all models are ferromagnetically ordered. As $p$ decreases, the strengths of both the bulk and the interfacial noise grow (starting from zero at $p=1$) in all the models. However, at any particular value of $p$, both sources of noise are \emph{stronger} in SCM3 compared to the weaker-than-soft version of the cooperative model. That is, with decreasing $p$, the effective noise grows faster in the SCM3 model. If a transition to the disordered regime is taken to happen once the noise in the system crosses a certain `threshold', then it is natural to expect that the onset of the paramagnetic phase would occur in SCM3 \emph{before} it occurs in the weaker-than-soft cooperative model, or, in other words, closer to the $p=1$ end point.

A similar comparison may also be made between the different versions of the cooperative model: the hard/soft rule on one hand and the weaker-than-soft rule on the other. All the three variants introduce identical \emph{interfacial} noise, because the outcome-based rules for conversion of the interfacial sites are the same for all of them. In addition, they all introduce the same \emph{transition probability} for the \emph{bulk} sites. However, in the weaker-than-soft case, the neighbors of the bulk site which is being updated are not persuaded to convert; as a result, updating of a bulk site would involve a conversion of \emph{only that} site (if at all). In contrast, when the hard version of the cooperative model is considered, the corresponding update rule for bulk sites involves not just the site which is currently being updated, but also a conversion of its \emph{neighbors} due to persuasion. In a sense, the conversion of a \emph{cluster} of sites in a sea (i.e., the neighbors \emph{plus} the updated bulk site) which happens in the hard version, is a bigger `fluctuation' than the conversion of \emph{just} a single site in a sea, which happens in the weaker-than-soft version; the probabilities in the two cases are, however, the same. If one then makes the assumption that the level of effective noise is `measured' not just in terms of how \emph{probable} changes are, but also by their `size', i.e., the \emph{number} of sites affected by update of a site, it would follow that the effective noise is uniformly stronger over the entire range of $p$ when persuasion is present. Thus, the hard version of the cooperative model may be anticipated to turn paramagnetic nearer to $p=1$ relative to its weaker-than-soft variant. The interpretation in the last line is lent support by recalling the analogy with the two-parameter generalized model (Drouffe and Godr\`{e}che 1999; Oliveira et al. 1993), which predicts a critical point at smaller value of interfacial noise (here, at larger $p$) when more bulk noise is present in the system. 


To conclude this section: the extension of the competitive learning model of Mehta and Luck (1999) to additionally include self-performance in the revising of the agents' individual choices is found to bring about some changes in the macroscopic dynamics of the population. The three choices for the update rule of bulk sites which were considered in this context introduce some systematic modifications, and these are reflected in the corresponding phase diagrams.  

\section{Discussion: Implications of our results for models of learning }

In this section, we will put our results in the context of existing models of learning, and analyze their implications.  Chatterjee and Xu (2004) have in a scholarly paper, mentioned several papers of relevance in the context of human behavior: models of cultural evolution (Bisin and Verdier 2001) which represent social conformity, and models of learning by
boundedly rational agents (Ellison and Fudenberg 1995,
 Eshel et al. 1998), which incorporate the tendency to learn from the successes of others. A realistic example of this can be found in agricultural innovation, where farmers employ fertilisers in response to their neighbors' successes (Conley and Udry 2000).  

Chatterjee and Xu (2004) build on these ideas to come up with
 two learning rules: one where the agent never switches type
 unless he fails, and one where he always looks around. In both cases the agent learns from the successes of his neighbours, as in the model of Mehta and Luck (1999). However in the model of Chatterjee and Xu (2004), there is no mechanism to change to a more successful type in a sea of failures, which exists both in
 the present model (SCM2 and SCM3), and in the cooperative model
of Mehta and Luck (1999). Another difference is that the analysis of Chatterjee and Xu (2004) was done not at coexistence, but where one of the probabilities of success was greater than the other; the dynamics of Ising spins makes it clear that in the general case, the magnetization will correspond to the sign of the dominant spin, i.e. the better technology will spread with probability 1, as was found in that paper. Analyzing the behavior at coexistence, as we have done, makes the generic behavior clearly visible in the phase diagrams presented here as well as in Mehta and Luck (1999). A third difference is that
the agents in the paper of Chatterjee and Xu (2004) make decisions based on immediate outcomes; in both this paper and that of Mehta and Luck (1999), we have set timescales such that an agent can make a decision based on the average outcome over a time period. It is easy to show (De Munshi et al. 2009) that the phase diagram depends critically on this choice; the conclusions of Bala and Goyal (1998) agree with this. Also, the stochastic selection of outcomes at any given timestep in our models is similar to that of Ellison and Fudenberg (1995).

To the best of our knowledge, none of the models mentioned above include the weighting of choices which include the agent's own performance, which is the main contribution of the present paper. In the case of the SCM1 model, the self-term acts to increase local order (the social conformity referred to above) by allowing the agent to switch types; while in the case of SCM2 and SCM3, the agent can, to varying extents, flip to a more successful type if it is in a sea of failures ('learning from one's failures'). This is midway between the inertia of earlier models in the case of failures (Chatterjee and Xu 2004) and the persuaded flipping of unsuccessful clusters via the hard and soft rules of the cooperative model of Mehta and Luck (1999). We view this as a rather realistic example of human behavior, which, when confronted by the failure of its own choices -- imposed by social conformity -- decides to change course individually.

So far, all our examples of learning were taken from cognitive/behavioral examples, which are in general those to which the present model is most applicable. There is however an interesting parallel to be found in the phenomenon of motor learning, where both learning and forgetting, as adaptive processes with different timescales have been shown to be important mechanisms in the phenomenon of reaching (Smith et al. 2006). In the original model of Mehta and Luck (1999), both learning and forgetting are part of the cognitive arsenal of a given agent, and it would be interesting to see, in future work, if the simple ideas of this paper could be used for modelling motor learning in humans and animals.

\section*{Appendix: Derivation of the probabilities $w_{\pm}(h)$}

The transition probabilities are calculated based on reasoning similar to that followed in arriving at the probabilities in Mehta and Luck (1999), but it is to be remembered that now the type and outcomes of the central site also play a role, and have to be included in calculating the ratios $r_+$ and $r_-$.

Let us begin with the derivation of the expression for $\bf w_+ (0)$: this is just the probability that $r_+ \geq r_-$, given that $\eta_i (t) = +1$ and $h_i=0$ (i.e. with a total of three `+' and two `$-$' sites \emph{including} the central site), and may be symbolically written as $P(r_+ \geq r_- \mid \eta_i = +1, h=0)$. Now, $P(r_+ = r_- \mid \eta_i = +1, h=0)$ is the probability that all sites succeed or all of them fail, which is just equal to $p^5 + (1-p)^5$. Next, to get $P(r_+ > r_- \mid \eta_i = +1, h=0)$, all pairs $(r_+,r_-)$ for which $r_+ > r_-$ have to be enumerated and their probabilities added up. The set of all such possible pairs is $\lbrace (1/3,0/2), (2/3,0/2), (2/3,1/2), (3/3,0/2), (3/3,1/2)\rbrace$, and their weighted sum is thus given by
\br
&&P(r_+ > r_- \mid \eta_i = +1, h=0) = \nn \\
&& 3p(1-p)^2 \cdot(1-p)^2 \nn \\
&& +~ 3(1-p)p^2 \cdot [(1-p)^2 + 2p(1-p)] \nn \\
&& +~  p^3 \cdot [(1-p)^2 + 2p(1-p)] \nn \\
&& = 3p(1-p)^4 + (3p^2 - 2p^3)(1-p^2), \nn
\er   
so $w_+ (0)$ finally reads
\br
w_+ (0) &=& P(r_+ \geq r_- \mid \eta_i = +1, h=0) \nn \\
&=& p^5 + (1-p)^5 + 3p(1-p)^4 \nn \\
&& +~ (3p^2 - 2p^3)(1-p^2).
\er
This expression can be rearranged to end up with the form given in Eq. (\ref{scm1}). All other $w$'s can be derived in a similar manner: \newline
$\bf w_+(+2)$: This is equal to $P(r_+ \geq r_- \mid \eta_i = +1, h = 2)$ with 4 `+' and 1 `$-$' sites, and is given by
\br
w_+(+2) &=& P(r_+ \geq r_- \mid \eta_i = +1, h = 2) \nn \\
&=& 1 - P(r_+ < r_- \mid \eta_i = +1, h = 2) \nn \\
&=& 1 - p \cdot (1-p^4) = 1 - p + p^5
\er
and the third line above makes use of the fact that $r_+ < r_-$ is possible only when the `$-$' type site succeeds and at least one `$+$' type site fails. \newline
$\bf w_-(+2)$: This is the probability that a `$-$' site will \emph{flip} its state, and is given by $P(r_- < r_+ \mid \eta_i = -1, h = 2)$ with 3 `+' and 2 `$-$' sites. The possible pairs $(r_+,r_-)$ for which this holds are $\lbrace(1/3,0/2),(2/3,0/2),(3/3,0/2),(2/3,1/2),(3/3,1/2)\rbrace$, and the total probability is thus
\br
w_-(+2) &=& P(r_- < r_+ \mid \eta_i = -1, h = 2) \nn \\
 &=& (1-p)^2 \cdot [1-(1-p)^3] \nn \\
& & + 2 \cdot p (1-p)\cdot[3p^2(1-p) + p^3] \nn \\
&=& (1-p)^2 - (1-p)^5 \nn \\
&& +~ 2p(1-p)(3p^2 - 2p^3)
\er
which can be brought into the form given in Eq. (\ref{scm1}). \newline
$\bf w_+(-2)$ and $\bf w_-(-2)$ can be directly obtained from the above probabilities: $w_+(-2) = 1- w_-(+2)$ and $w_-(-2) = 1-w_+(+2)$. \newline
$\bf w_+(+4)$($\bf w_-(-4)$) would be equal to 1(0) since the central site has no way to make a comparison between the outcomes of the two types. On the other hand, if $\eta_i (t) = -1$ and $h= +4$, the central site can make such a comparison by considering its own outcome too; it will flip to the other type if it fails \emph{and} at least one of its neighbors (all of which are `+') succeeds, so
\br 
w_-(+4) &=& P(r_- < r_+ \mid \eta_i = -1, h = 4) \nn \\
&=& (1-p)\cdot[1-(1-p)^4] \nn \\
&=& 1 - p - (1-p)^5
\er
and this consequently gives ${\bf w_+(-4)}=1-w_-(+4)=p+(1-p)^5$.

\begin{acknowledgements}
G.M. would like to thank Nirmal Thyagu for helpful discussions in the course of this work. G.M. was supported by a grant from DST (Govt. of India) through the project ``Generativity in Cognitive Networks''.
\end{acknowledgements}


\begin{thebibliography}{99}

\bibitem{bg}
Bala V, Goyal S (1998) Learning from Neighbours. Review of Economic Studies,  65: 595-622

\bibitem{stripes1}
Barros K, Krapvisky PL, Redner S (2009) Freezing into stripe states in two-dimensional ferromagnets and crossing probabilities in critical percolation. Phys Rev E 80 040401

\bibitem{bisin}
Bisin A, Verdier T (2001) The Economics of Cultural
Transmission and the Dynamics of Preferences. Journal of Economic Theory, 97: 298-319

\bibitem{socdyn}
Castellano C, Fortunato S, Loreto V (2009) Statistical Physics of Social Dynamics. Rev Mod Phys 81: 591-646

\bibitem{kc}
Chatterjee K, Xu SH (2004) Technology diffusion by learning from neighbours. Adv in Appl Probab: 355-376

\bibitem{fss} 
Christensen K, Moloney NR (2005) Complexity and Criticality. Imperial College Press, London

\bibitem{conley}
Conley TG, Udry CR (2000) Learning about
a New Technology: Pineapple Growers in Ghana. Mimeo Northwestern
and Yale Universities

\bibitem{dtm}
De Munshi D, Thyagu NN, Mehta A (2009) Unpublished

\bibitem{voter1}
Dornic I, Chat\'{e} H, Chave J, Hinrichsen H (2001) Critical Coarsening without Surface Tension: The Universality Class of the Voter Model. Phys Rev Lett 87, 4 045701: 1-4

\bibitem{gim1}
Drouffe JM, Godr\`{e}che C (1999) Phase ordering and persistence in a class of stochastic processes interpolating between the Ising and voter models. J Phys A: Math Gen 32: 249-261

\bibitem{ellison}
Ellison G, Fudenberg D (1995) Word-of-mouth Communication
and Social Learning. Quarterly Journal of Economics, 110: 93-125

\bibitem{sam}
Eshel I, Samuelson L, Shaked A (1998) Altruists, Egoists
and Hooligans in a Local Interaction Model. American Economic
Review 88: 157-179

\bibitem{voter2}
Frachebourg L, Krapivsky PL (1996) Exact results for kinetics of catalytic reactions. Phys Rev E 53: R3009-R3012

\bibitem{rs-updating1}
Glauber RJ (1963) Time-dependent statistics of the Ising model. J Math Phys 4: 294-307

\bibitem{voter3}
Holley R, Liggett T (1975) Ergodic theorems for weakly interacting infinite systems and the voter model. Ann Probab 3: 643-663
 
\bibitem{rs-updating2}
Landau D, Binder K (2005) A Guide to Monte Carlo
 Simulations in Statistical Physics. Cambridge University
 Press, Cambridge, UK

\bibitem{clm}
Mehta A, Luck JM (1999) Models of competitive learning: Complex dynamics, intermittent conversions,
and oscillatory coarsening. Phys Rev E 60, 5: 5218-5230

\bibitem{gim2}
Oliveira MJd, Mendes JFF, Santos MA (1993) Nonequilibrium spin models with Ising universal behaviour. J Phys A: Math Gen 26: 2317-2324

\bibitem{motorlearning}
Smith MA, Ghazizadeh A, Shadmehr R (2006) Interacting adaptive processes with different timescales underlie short-term motor learning. PLoS Biology, 4: e179

\bibitem{stripes2}
Spirin V, Krapivsky PL, Redner S (2001a) Freezing in Ising ferromagnets. Phys Rev E 65 016119: 1-9

\bibitem{stripes3}
Spirin V, Krapivsky PL, Redner S (2001b) Fate of zero-temperature Ising ferromagnets. Phys Rev E 63 036118: 1-4
 
\end{thebibliography}
\end{document}